\documentclass[conference]{IEEEtran}
\usepackage[pdftex]{graphicx}
\usepackage{cite}
\usepackage{url}
\begin{document}
\thispagestyle{plain} % for page numbers
\pagestyle{plain}     % for page numbers
\title{DDTS: A Practical System Testing Framework\\for Scientific Software}
\author{\IEEEauthorblockN{Paul Madden}
\IEEEauthorblockA{NOAA Earth System Research Laboratory\\
University of Colorado at Boulder \\
Boulder, Colorado\\
paul.a.madden@noaa.gov}
\and
\IEEEauthorblockN{Eduardo G. Valente Jr.}
\IEEEauthorblockA{NASA GSFC Earth Sciences Division\\
Global Science and Technology Inc.\\
Greenbelt, Maryland\\
eduardo.g.valente@nasa.gov}}
\maketitle
\begin{abstract}
Many scientific-software projects test their codes inadequately, or not at all. Despite its well-known benefits, adopting routine testing is often not easy. Development teams may have doubts about establishing effective test procedures, writing test software, or handling the ever-growing complexity of test cases. They may need to run (and test) on restrictive HPC platforms. They almost certainly face time and budget pressures that can keep testing languishing near the bottom of their to-do lists. This paper presents DDTS, a framework for building test suite applications, designed to fit scientific-software projects' requirements. DDTS aims to simplify introduction of rigorous testing, and to ease growing pains as needs mature. It decomposes the testing problem into practical, intuitive phases; makes configuration and extension easy; is portable and suitable to HPC platforms; and exploits parallelism. DDTS is currently used for automated regression and developer pre-commit testing for several scientific-software projects with disparate testing requirements.
\end{abstract}
\section{Introduction}
Professional software engineers recognize routine testing as an indispensable best practice \cite{prags}\cite{fowler-ci}\cite{extreme}\cite{batlab}, on par with the use of revision-control systems. Testing provides a bulwark against regression, makes refactoring and optimization tractable, and helps assure reproducible results. The risk of undesirable outcomes -- from large chunks of time sunk in debugging \cite{comp-sci-survey} to portability issues \cite{cig} to retracted papers \cite{does-not-compute} -- can be mitigated, and huge software-development investments protected, through a commitment to systematic software testing.

But, while advocated for on behalf of the scientific-software community \cite{bp4sc}, and accepted in principle by the community itself, testing is often embraced late, or forgone entirely, for a variety of reasons. Scientists who develop software rate their understanding of testing concepts lower than they do the importance of testing itself, and may harbor (legitimate) doubts related to the usefulness of testing in the face of numerical approximation errors, difficulties explicitly stating requirements, and the exploratory nature of scientific software \cite{how-do}. They may be unaware of the benefits of testing, or of existing techniques they might adopt \cite{bottleneck}. Some may assume that adequate testing is done by the authors of software components they reuse, or believe that their code is too simple to require testing \cite{comp-sci-survey}.

This paper presents a system testing framework, DDTS (for \emph{Dependency-Driven Test System}), that seeks to ease the introduction of routine testing into scientific-software development efforts. In line with the goals of organizations like Software Carpentry \cite{sw-carp} and The Software Sustainability Institute \cite{ssi}, DDTS is concerned with helping development teams adopt testing sooner rather than later, via a relatively simple tool appropriate to their needs, or at least to the needs of a typical scientific-software project. Since small teams may consider testing less important than larger ones \cite{how-do}, and since they likely have fewer resources at their disposal to begin with, a low barrier to entry may make the difference between the adoption of testing and its neglect -- an important consideration given that small projects may grow into large ones.

DDTS exposes a small set of stub software routines which, when fleshed out to describe the build, run and other activities pertinent to the program-under-test, specify an interface between the framework and that program, and provide a somewhat opinionated recipe for testing. These routines may call (in shells spawned by the framework) existing build- and run-automation scripts developers already run by hand, and so include these utilities, whose continued correct operation is certainly worthy of routine testing, under the system testing umbrella.

This paper's contributions are the description of a set of requirements, derived from and validated through experience, for a practical system testing framework that can be easily applied to a variety of scientific-software codes; and a discussion of how these requirements are realized in DDTS. As evidence of the framework's general applicability -- and, it is hoped, its suitability for use by fledgling or even established software projects -- we describe test-suite applications created for several atmosphere/climate model codes at both the U.S. National Oceanic and Atmospheric Administration (NOAA) and the U.S. National Aeronautics and Space Administration (NASA). We generally omit details on internal mechanisms, use of design patterns, etc. and instead describe those implementation details that benefit developers using DDTS to implement a system test suite for their software project.

\section{Related Work}

DDTS follows a \emph{system testing} approach. Unlike unit testing, which focuses on software's basic units -- functions and subroutines -- system testing is concerned with the behavior of the program-under-test as a whole. While unit testing is a powerful and desirable technique, it can in practice be difficult to apply to legacy science codes \cite{clune} (e.g. due to programming habits common in languages like Fortran, like long subroutines that mutate global data), and in cases where it is difficult to establish ``passing'' criteria for floating-point implementations of some complex algorithms. The detailed knowledge of a routine required to write an effective unit test may only be available to domain experts, limiting collaboration with other developers (e.g. software engineers) in building up test suites.

Acknowledging that the perfect need not be the enemy of the good, system testing avoids some of these difficulties by testing software at a higher level of abstraction, and can be an appropriate starting place for many projects. Those not already using unit testing can easily obtain some quality assurance by adopting system testing via a framework like DDTS. Projects already taking advantage of a unit testing framework like JUnit \cite{junit} or pFUnit \cite{pfunit} can achieve even higher levels of assurance through the addition of system tests, which may also test external components like file, batch and database systems.

The remainder of this paper discusses DDTS' design criteria and implementation; current extensions to the original design; experiences applying DDTS to specific scientific codes; and possible future work.

\section{Design Criteria}
This section discusses the several criteria that guided the design of DDTS.

\subsection{Design criterion: The framework should organize test activities using the intuitive concepts of builds, runs, groups and suites.}

\begin{figure}[!t]
\centering
\includegraphics[width=3in]{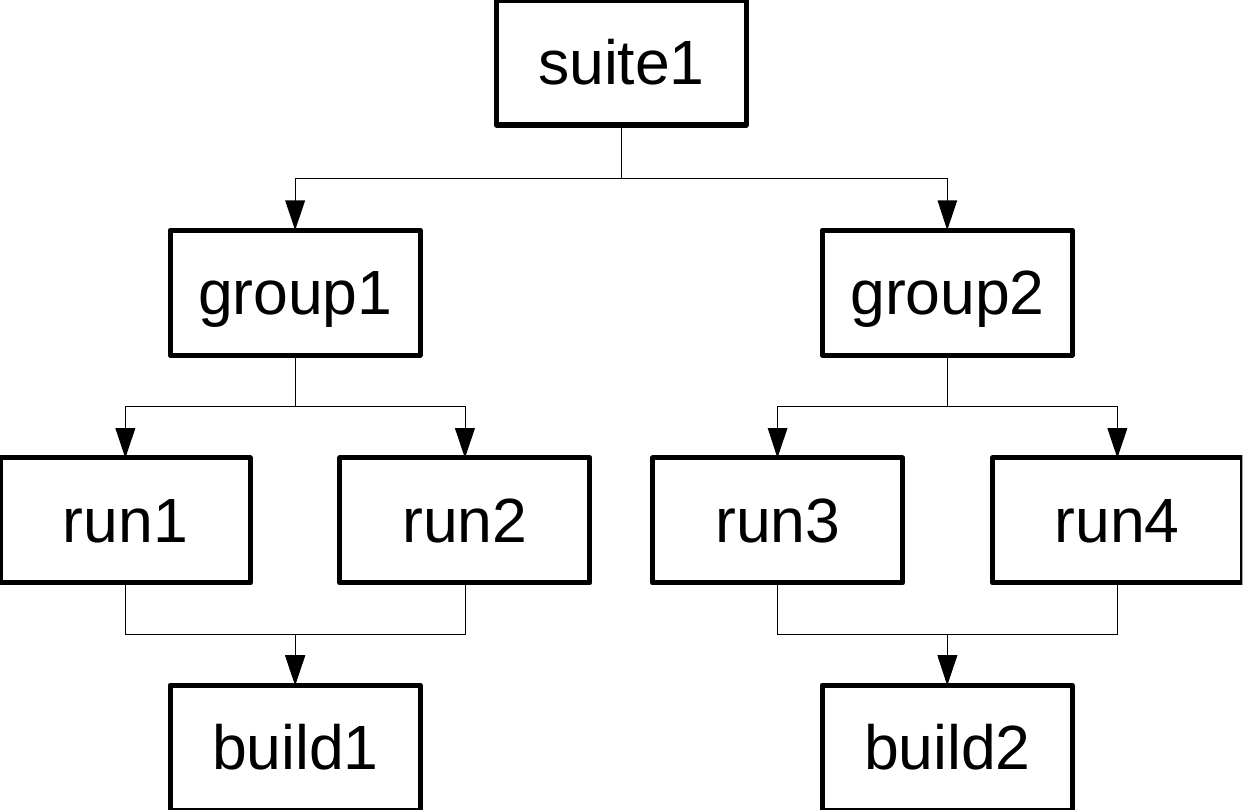}
\caption{Test activities can be represented in terms of \emph{suites}, comprising a number of comparison \emph{groups}; \emph{runs}, whose membership in a common group declares the expected equivalence of their output; and \emph{builds}, which produce the executable programs required by runs. As shown, multiple runs may depend on the same underlying build.}
\label{figure:1}
\end{figure}

A practical conceptual structure for the organization of test activities employs the ideas of \emph{builds}, \emph{runs}, \emph{groups} of runs, and \emph{suites} of groups. The build and run concepts map onto their obvious counterparts outside the framework: builds onto the executable programs and libraries created from source code, and runs onto individual executions, with specific configurations, of those programs. Groups represent sets of comparisons between runs' output, and exist to tie together runs expected to produce output that is ``equivalent'', in a sense described later. Suites represent collections of these comparison groups. Figure \ref{figure:1} shows an example hierarchy of builds, runs, groups under a suite.

\subsection{Design criterion: Run-vs-run and run-vs-baseline comparison methods should be available.}

Two useful methods for judging the correctness of an execution of the program-under-test are \emph{run-vs-run} and \emph{run-vs-baseline} comparisons. The framework should support both.

Throughout this paper, \emph{run} refers to a single execution of the program-under-test, and should not be confused with the execution of the testing framework itself, referred to in this paper as a \emph{test-suite invocation}. A single test-suite invocation may perform many runs, perhaps representing various configurations of the program-under-test.

\subsubsection{Run-vs-run comparisons}

In a run-vs-run comparison, one run's output is compared to that of another (with a different but compatible configuration) executed in the same test-suite invocation, with the expectation that the two will be equivalent.

Here, ``equivalent'' simply means that some defined comparator function (or \emph{oracle}, e.g. ``bitwise identical'') judges both sets of output to represent the same results.

``Compatible'' means that any configuration differences are not expected to change the behavior of the program in a way that would affect output equivalence. For example, two runs might use different numbers of MPI (Message Passing Interface) tasks or OpenMP threads, or different parallel IO mechanisms, and these differences would not be expected to lead to different output.

Run-vs-run comparisons are useful for detecting cases where programming errors lead to unwanted differences (e.g. due to using data from uninitialized arrays), and are especially helpful for detecting parallelization errors (e.g. using data from out-of-date halo regions). It is the standard generally adhered to when domain experts modify a scientific code: Their work is \emph{expected} to change (and hopefully improve) a program's output -- but not to introduce inconsistencies between compatible runs.

\subsubsection{Run-vs-baseline comparisons}

In a run-vs-baseline comparison, a run's output is tested for equivalence, in the sense described above, to that of a corresponding run executed by an earlier test-suite invocation and stored as part of a ``baseline'' collection of output from one or more runs. This method ensures against unexpected changes in the tested program's output. It is the standard generally adhered to by software engineers working on a scientific code, and ensures that refactoring and optimizations do not change results. The run-vs-baseline method helps domain experts to avoid unintended changes output.

Note that neither of these methods relate to output validation/verification (correctness of the output's \emph{meaning}), which must be established by other means.

\subsection{Design criterion: Tests should be simple to run, and the summary report easy to interpret.}

To support regular, consistent use of test suites as part of the development cycle, the framework should provide a simple means to execute test suites, with unambiguous pass/fail result messages. A test suite may execute any number of individual runs, each perhaps requiring the compilation of executable programs, provisioning of input data, creation of run-time directories, etc., as well as some number of comparisons to establish suite success or failure. End users (people simply running a test suite) benefit from the automation of these tedious tasks, and can be shielded from the details of their execution -- at least until tests fail. The potentially voluminous information collected during execution of a test suite should be retained, as it may be useful for post-mortem analysis and debugging.

A balance should be maintained between terse progress information shown to the user, and verbose information collected in a log file.

\subsection{Design criterion: The framework should be lightweight and portable.}

Many scientific-software development efforts are pursued on shared platforms (computer systems) where developers may not have permission to install new system software, establish network services, or communicate freely with remote systems. This is commonly the case on high-performance computing (HPC) platforms. The test framework should therefore be lightweight: It should rely on as few external components (e.g. web servers, databases) as possible, and should provide a command-line interface to support operation on restrictive platforms. A need to convince system-administration staff to install extra software components to support testing should be avoided.

It may also be necessary to develop and test on multiple platforms. While the program-under-test might not change when moving to a new platform, resources like compilers, libraries and batch systems likely will. Some of this diversity may be hidden by the program's existing build- and run-automation systems (scripts, makefiles, etc.). The framework should make it simple to define test suites for each platform, reusing existing test-suite code and build/run automation utilities wherever possible. That is, once the interface between the framework, the program, and one platform is established, it should be straightforward to redefine pieces of this interface to support new platforms.

\subsection{Design criterion: Configuration, extension and maintenance should be simple.}

As with the requirement for easy portability of a test suite to multiple platforms, the framework should similarly make it easy to extend the definitions of suites, runs and builds. The guiding software-engineering principle here is DRY (\emph{Don't Repeat Yourself}) \cite{prags}: If a new build, run or suite definition is needed, and if an existing one is close to what is required, it should be possible to automatically produce the new definition by composing the old one with a set of desired modifications. This should be done automatically by the framework each time a test suite is invoked, replacing error-prone, duplicative copy-and-paste with DRY dynamic derivation. The framework should therefore provide an inheritance mechanism to support extensibility, similar to the inheritance mechanisms of object-oriented programming languages.

To encourage domain experts to compose and modify tests, it should be simple to define builds, runs and suites without learning a new programming language or being exposed to framework internals. Two benefits of a \emph{declarative} approach recommend it for this task: First, its simplicity welcomes a wider audience of users than a traditional programming language might; and second, unlike an \emph{imperative} syntax that would imply an order of execution, opportunities for parallelism can be automatically extracted from declarative forms \cite{decl-advantages}. A declarative syntax should therefore be used to define those high-level test-suite activities that can be executed concurrently.

Finally, developers defining or running test suites should not be forced to consider the order in which builds, runs and comparisons must be carried out, the actual mechanics of their execution, or how information might be shared between them. Given the existence in the test-suite application of appropriately implemented interfaces between the framework, program and platform, the framework should deduce from the declarative definitions the dependencies between them, and so the order in which they must be executed, and should arrange for information sharing between them. This minimizes maintenance costs as test suites grow in size and complexity.

\subsection{Design criterion: Parallelism should be used for reasonable turnaround times.}

The configuration space of a scientific code is often huge, and system tests typically can cover only a small fraction. Even then, a test suite providing reasonable coverage may require dozens or even hundreds of builds and runs. To achieve practical turnaround times (crucial when a test-suite pass is a prerequisite for a commit to a revision-control repository), the framework should exploit parallelism to overlap execution of independent components when the underlying platform supports efficient concurrent execution.

\section{Implementation}

This structure of this section mirrors that of the previous one to discuss how each of these design criteria is currently realized in DDTS.

\subsection{Implementation: The framework should organize test activities using the intuitive concepts of builds, runs, groups and suites.}

Simple text files defining builds, runs, groups and suites control the high-level testing activities performed by DDTS, and are kept in subdirectories appropriately named \emph{builds}, \emph{runs} and \emph{suites}. The definitions contained in these files are referred to in other definition files by their pathless filenames. For example, a build definition contained in the file \emph{builds/gfortran} would be referenced in a run definition file simply as \emph{gfortran}. These references are used to build up suite definitions from run definitions, and to tie runs to builds.

A typical DDTS invocation of a certain named suite would start by determining its prerequisite \emph{group} components, then the \emph{run} prerequisites of each group, etc., leading to the eventual execution of all activities in the required order. This system is described further in a later subsection.

\subsection{Implementation: Run-vs-run and run-vs-baseline comparison methods should be available.}

In DDTS, a suite definition is composed of one or more named groups, each of which contains the names of one or more runs (actual syntax is shown in a later subsection). The presence of multiple runs in the same group directs DDTS to verify the equivalence of their output as a condition of test-suite success. This concise notation not only declares that the named runs must be (successfully) executed, but also implicitly requests performance of run-vs-run comparisons of their respective output files. No special command-line syntax is needed when invoking DDTS to obtain this essential behavior.

DDTS' \emph{use-baseline} command-line option requests that run-vs-baseline comparisons also be performed when executing a test suite (or even a single run). The option's argument is a path to a directory containing a set of baseline output generated by a previous test-suite invocation. Any DDTS run's definition may include an optional \emph{baseline} key associating it with a specific named baseline, and the framework will compare the output of each run so defined with the matching set of baseline output in the specified directory.

Similarly, the \emph{gen-baseline} command-line option requests that a baseline be generated from the output of a suite (or single run). Its argument names a directory where the baseline should be stored. Since more than one run in a test-suite invocation may associate with the same baseline, only one should contribute its output to the baseline. The mechanism governing this is described in a later subsection.

By default, DDTS uses a bitwise-exact comparator to judge whether output files are equivalent, but custom comparators may be specified in suite and run definition files. This is discussed further in a later subsection.

Note that, because runs need not associate with \emph{any} baseline, and because groups defined in suite definitions are allowed to contain only a single run, it is possible to define ``smoke test'' runs expected merely to run to completion, and whose output is not tested at all. These runs may, in fact, not exercise the primary program-under-test, but buttress the test suite's overall assurances by testing related code or even platform components.

\subsection{Implementation: Tests should be simple to run, and the summary report easy to interpret.}

In the most common use case, DDTS can be invoked with a single command-line argument: the name of a defined test suite to execute (e.g. \emph{ddts suite1}). As discussed above, a suite definition (usually) implies a set of run-vs-run comparisons, described as a comparison group, that test for equivalence of output within the group and for the ability to successfully build, configure and run the program-under-test to completion.

Execution of a single run (e.g. \emph{ddts run run1}) is also supported, and useful, for example, when iterating on a failed test, evaluating possible fixes, before executing the entire suite again.

DDTS limits screen output, so as not to overwhelm the user with information. During execution, it prints basic progress messages when, for example, builds or runs start and finish, comparisons are performed, baselines are created, and in several other informational or warning cases. It reports errors as they occur and, if none are detected, prints a final definitive ``ALL TESTS PASSED'' message.

Significantly more verbose output is collected in a unique log file created for each DDTS invocation. For example, the results of individual output file comparisons are logged, as is the entire output generated by external shell commands executed by the framework. All messages printed to the screen are also logged, so that the log file is a complete record of each DDTS invocation's activities.

In case specific applications require more screen or log file output than is generated by default, DDTS provides routines to write to both the screen and the log file, callable by the code written by test-suite application implementers to interface DDTS to the platform and program-under-test.

\subsection{Implementation: The framework should be lightweight and portable.}

Compute platforms and resource allocations come and go, and a scientific-software development effort may outlive both, making portability of a code -- as well as its tests -- valuable. Portability also benefits teams wanting to run their code on multiple platforms at once, for example to take advantage of a variety of hardware and software configurations. In recognition of these needs, DDTS' implementation supports simple portability.

DDTS is written in Ruby \cite{ruby} and designed to run on the JRuby \cite{jruby} implementation which, in turn, runs on the Java Virtual Machine.

Ruby is a high-level, object-oriented, dynamically-typed, interpreted language with a concise syntax and a robust standard library -- features that were all useful in the development of the framework. Of course, many languages offer similar benefits, and Java itself is famously portable. More important, then, is the advantage Ruby confers on the developers of library routines (see below), which may be written in a boilerplate-free scripting style likely to be comfortable for many would-be DDTS application implementers who may have experience writing shell scripts, Perl, Python or, of course, Ruby.

The decision to target JRuby was based on the observation that finding or provisioning a reasonably modern Java (JRuby's only external requirement) on any given platform -- especially an HPC platform -- is easier than finding an up-to-date native Ruby installation. Also, since JRuby emulates several Ruby versions, DDTS can support only one and maintain portability via JRuby's support for that version.

DDTS is command-line based, requires only Java and a single JRuby jar file, and writes only to the screen and to its log files, so that it requires no database, web server, or other external service. It is intended to interact with the program-under-test primarily by automating actions users would normally execute by hand.

Before discussing portability via specialization of the platform interface, some background on a DDTS application's ``library'' routines is necessary. The purpose of this detour is to illustrate what the task of applying DDTS to a to-be-tested piece of software involves: which routines can be defined, what they do, how information from YAML definition files is made available to the library routines, etc.

\subsubsection{DDTS Library Routines}

\begin{table}[!t]
\centering
\begin{tabular}{r c c c c c c}
\hline
Order & Routine & Called From & Purpose \\ 
\hline
1 & lib\_suite\_prep & test-suite core thread & suite-wide setup \\
2 & lib\_build\_prep & run thread & pre-build setup \\
3 & lib\_build & run thread & main build actions \\
4 & lib\_build\_post & run thread & post-build teardown \\
5 & lib\_data & run thread & prepare data for all runs \\
6 & lib\_run\_prep & run thread & pre-run setup \\
7 & lib\_run & run thread & main run actions \\
8 & lib\_run\_post & run thread & post-run teardown \\
9 & lib\_run\_check & run thread & check for run success \\
10 & lib\_outfiles & run thread & identify the run's output files \\
11 & lib\_comp & run thread & for run-vs-baseline comparison \\
12 & lib\_comp & comparison-group thread & for run-vs-run comparison \\
13 & lib\_suite\_post & test-suite core thread & suite-wide teardown \\
\hline
\newline
\end{tabular}
Additionally, lib\_queue\_del\_cmd is called asynchronously, by each run thread, if the test suite fails and a batch system is in use.
\newline
\caption{DDTS library routines. }
\label{table:routines}
\end{table}

The interface between DDTS, the program-under-test and the execution platform is defined by thirteen routines (Ruby methods). A source file containing correct but useless stub versions of these routines is packaged with DDTS as \emph{defaults.rb}. An application of DDTS to the program-under-test and execution platform requires that some or all of these routines (depending on the needs of the program and platform) be overridden, in the file \emph{library.rb}, with versions whose bodies do actual work.

Six of the routines are responsible for performing prep (e.g. setup), main (e.g. test-case execution) and post (e.g. teardown) work related to executing builds (\emph{lib\_build\_prep}, \emph{lib\_build} and \emph{lib\_build\_post}) and runs (\emph{lib\_run\_prep}, \emph{lib\_run} and \emph{lib\_run\_post}). This division is arbitrary, and left intentionally undefined, but experience indicates that this breakdown is useful for organizing build and run activities, and that many fall naturally into these categories. Ultimately, it is up to the library implementer to decide how activities should be grouped together. For some test-suite applications, the authors have found it useful to do meaningful work in each of the routines; in others, some routines are simply pass-through stubs.

Two routines, \emph{lib\_suite\_prep} and \emph{lib\_suite\_post}, provide a place to execute general setup and teardown activities on behalf of the entire test suite, before any and after all builds or runs are performed.

The remaining five routines are responsible for actions like returning the list of output files created by the program-under-test that are subject to comparison; making input data needed by test-suite runs available prior to execution of those runs; determining whether or not a single run completed successfully (independent of the quality of its output); removing a queued or running job from the batch system, if one is in use, when a test suite fails; and providing an alternative to the default bitwise-exact comparator.

Table \ref{table:routines} lists the DDTS library routines, and the order in and context from which they are called. It is evident that most work is done by run threads. As discussed later, when several runs depend on a common build, the three \emph{lib\_build*} routines will be executed by a single run thread, and the resulting executable(s) shared by those runs.

The \emph{lib\_comp} routine is potentially called twice: Once, when a run thread compares its output to its baseline, and again when a comparison group compares the output of two runs against one another.

Note that, excepting the two \emph{lib\_suite\_*} routines, the order in which the library routines are called is determined by the recursive, bottom-up, dependency-driven nature of the test-suite definitions: For example, builds must be performed before runs, output is identified after runs complete, and comparisons are performed based on the identified output.

\subsubsection{Library Routine Arguments and Return Values}

DDTS' core driver calls library routines directly, so their argument lists and expected return values are defined and documented. Some arguments are flexibly defined, with details left to the discretion of the library implementer. For example, \emph{lib\_run} is allowed to return an arbitrary object that will later be passed to \emph{lib\_run\_post}. The library implementer, knowing from the documentation that \emph{lib\_run\_post} will be responsible for determining whether or not the run completed successfully, should determine the specific type and value of the returned object. On the other hand, the structure and contents of the return value required from \emph{lib\_outfiles} is precisely defined in the DDTS documentation.

\subsubsection{The \emph{env} argument}
The first argument to each of the library routines is an object called \emph{env}, which contains a structure initially reflecting only the build, run and suite definitions pertaining to this particular DDTS invocation, and appropriate to each of the test-suite execution phases. For example, in a library routine called from the context of a run, the \emph{env.run} substructure will contain keys and values from that run's definition file (e.g. \emph{env.run.build} will contain the name of the build used by this run, and \emph{env.run.baseline}, if defined, will contain the name of the baseline with which this run is associated). The syntax of the build, run and suite definitions are discussed later, but it is useful to note here that the \emph{env} structure may be arbitrarily modified by library routines, according to the implementer's needs. In particular, it may be useful to pass information between library routines via the \emph{env} structure, beyond what is required or allowed by the routines' other arguments and return values.

Returning now to the topic of portability, a particular DDTS application might require that, for example, the \emph{lib\_run} routine use a certain batch-submission command on one platform, and a different command on another. DDTS supports this requirement via the build, run and suite definition files: If a library routine name appears as a key in a definition file, that key's associated value names a routine to be called in place of the original. For example, if a run's definition file contains the key-value pair \emph{lib\_run}: \emph{lib\_run\_sge}, then the DDTS core driver will invoke \emph{lib\_run\_sge} for this run where it would normally have invoked \emph{lib\_run}. The \emph{lib\_run\_sge} routine must then be defined in \emph{library.rb} and would execute a job-submission command appropriate to the Sun Grid Engine (SGE) batch system. A run defined for a different platform might specify \emph{lib\_run}: \emph{lib\_run\_pbs} to submit a job to the Portable Batch System (PBS). This mechanism allows specialization of library routines for platform portability. Ruby, as a dynamic language, makes this aliasing feature trivial to implement.

\subsection{Implementation: Configuration, extension and maintenance should be simple.}

A full DDTS application involves three distinct types of programming tasks, suited to three programmer skill levels, a division that leads to a useful separation of concerns.

First, and perhaps most complex, is the DDTS core driver, which handles the overall test-suite workflow, including thread and mutex management, logging, command-line and definition-file processing and baseline generation and comparison. It is written and maintained by the DDTS developers and is not intended to be modified for any particular test-suite application.

Second, and somewhat less complex, is the library implementation, where the program and platform interface routines are defined. The complexity of these routines depends on an application's needs, but tends to be low. They often resemble simple shell scripts that mimic actions users would manually perform on the command line when building software, configuring it, submitting batch jobs, etc. Still, the library must be implemented by someone familiar with both the program-under-test and with DDTS itself, and able to perform basic Ruby programming. Once implemented, however, the program and platform interface defined in the library usually requires little maintenance.

Third, by design least complex, and intended for modification by anyone who might need to create or execute test suites, are the declarative definition files that control high-level test-suite behavior.

Two advantages of a declarative approach are, first, that it simplifies programming; and, second, that it is agnostic about execution order and so encourages parallelism. Both advantages can be explained in terms of Kowalski's \emph{algorithm = logic + control} concept \cite{logic-control}: A program's declarative logic component describes the problem to be solved, while its imperative control component prescribes the actual mechanics of execution. Different control components can be used to solve the same logically defined problem, and the logical declaration language can remain unaware of control details.

\begin{figure}[!t]
{\small \begin{verbatim}
$ ddts show suite zeus

group_zeus_intel: 
  zeus_intel_mpt_1
  zeus_intel_mpt_10
  zeus_intel_mpt_20
  zeus_intel_serial
group_zeus_lahey: 
  zeus_lahey_serial
\end{verbatim} }
\caption{A request to show a YAML suite definition, displaying two comparison groups, \emph{group\_zeus\_intel} and \emph{group\_zeus\_lahey}. The output of the four runs in \emph{group\_zeus\_intel} will be automatically inter-compared by the framework. Since the run \emph{zeus\_lahey\_serial} is alone in its group, its output can only be compared to a baseline, if the run is associated with one. The suite is considered successful if both comparison groups are successful, which requires that the output of the runs under each group are deemed equivalent by the defined comparator function.}
\label{figure:2}
\end{figure}

DDTS' logic component consists of a set of build, run and suite definition files, expressed in YAML \cite{yaml}. They provide, in a sense, simple domain-specific languages. A suite definition (Figure \ref{figure:2}), for example, poses the question ``Is the output produced by each run in each of the defined comparison groups equivalent to the output of the other runs in the same group?'' An affirmative answer implies a test-suite pass. The mechanics of how this question is answered -- that is, how a run is performed, where its output is located, how output files are compared, etc. -- are not specified at this level. Nor is the order in which runs must be performed made explicit, which admits the possibility that they might execute in parallel. These details are made explicit in the DDTS library defined for the application, in cooperation with the framework's core driver.

The simplicity of YAML, examples of which can be seen in Figures \ref{figure:2} and \ref{figure:3}, owes much to its light use of punctuation and reliance on indentation to express structure. This makes reading and writing definitions easy, and helps scientific-software developers to define their own suites by creating groups and assigning runs to them. They need to understand almost nothing about DDTS or the library implementation to create complex suites comprising builds, runs, comparisons, etc.

DDTS supports very flexible run definitions. Except for the required \emph{build} and optional \emph{baseline} keys, the contents of a run definition depend only on what information the library implementer decides is required by the run-related library routines, and is made available to these via the passed-in \emph{env} object. But, generally speaking, a run definition poses the question ``If a run of the program-under-test, configured with these parameters, successfully runs to completion, what is its output and where can it be found?'' This is the information required by the comparison group, one level up in the execution hierarchy, to determine its success or failure.

\begin{figure}[!t]
{\small \begin{verbatim}
$ ddts show run jet_pgi_cpu_nophys_10

# jet_pgi_cpu_nophys_10 < jet_std < nim_base

baseline: jet_pgi_nophys
build: jet_pgi_cpu_p
lib_build_prep: lib_build_prep_jet
lib_run_prep: lib_run_prep_std
namelist_file: NIMnamelist
namelists:
  cntlnamelist:
    glvl: 5
    nz: 32
    physics: none
  computetasknamelist:
    computetasks: 10
  queuenamelist:
    maxqueuetime: 00:20:00
\end{verbatim} }
\caption{A request to show a run definition. The ancestry line, prefixed by \#, shows that the run was defined as an extension of run \emph{jet\_std}, which in turn extends \emph{nim\_base}. (See Figure \ref{figure:4} for an example of how definition information is actually reused via \emph{extends}.) The (required) \emph{build} and (optional) \emph{baseline} keys are defined here, and two library routines, \emph{lib\_build\_prep} and \emph{lib\_run\_prep}, are associated with specialized implementations. Since this is a run definition, the value of \emph{nz} (for example) would be obtained in a run-related library routine by referencing \emph{env.run.namelists.cntlnamelist.nz}.}
\label{figure:3}
\end{figure}

To honor the DRY principle and support code reuse, DDTS build, run and suite definition files support an \emph{extends} key, whose value is the name of another definition, of the same type, to use as a base to build upon. This mechanism resembles single inheritance in object-oriented design. Final definitions are composed recursively, starting with each definition's \emph{extends} ancestor, then updating it dynamically with the keys/values from the current definition. Cycles in \emph{extends} ancestry chains are detected and treated as errors. DDTS provides a \emph{show} command (e.g. \emph{ddts show run run1}) that displays the fully-composed definition for a given build, run or suite, along with its ancestry. See Figure \ref{figure:3} for an example.

\begin{figure}[!t]
{\small \begin{verbatim}
$ cat runs/run3
build: build2
key1: value1

$ cat runs/run4
extends: run3
key2: value2

$ ddts show run run4

# run4 < run3

build: build2
key1: value1
key2: value2
\end{verbatim} }
\caption{The \emph{extends} keyword defines a new run based on another. Here, a new run, \emph{run4}, extends \emph{run3}, then adds a new key-value pair, \emph{key2: value2}. When the definition of \emph{run4} is displayed by DDTS' \emph{show} command, the composed result, as well as its ancestry, are shown.}
\label{figure:4}
\end{figure}

\begin{figure}[!t]
{\small \begin{verbatim}
$ cat suites/suite1
group1:
  - run1
  - run2
group2:
  - run3
  - run4 # newly-added run
\end{verbatim} }
\caption{Once a new run definition file, e.g. \emph{runs/run4} is created, adding it to an existing comparison group (\emph{group2} here) triggers automatic execution of the run by the framework, and comparison of its output to that of the other runs in the group, when the suite is executed.}
\label{figure:5}
\end{figure}

Using the \emph{extends} mechanism, developers can, for example, define a new run by extending the definition of an existing run and simply overriding definition parameters of interest, or adding new parameters (Figure \ref{figure:4}). Since DDTS definition names correspond to filenames, once a new run definition file is created in the appropriate directory, it can be incorporated into a test suite by adding its pathless filename to the list of runs in an existing (or new) comparison group. When that suite is next executed, the new run will be performed and its output compared against any other runs in its group. Figure \ref{figure:5} shows an example of a suite definition file with a newly-added run. (The \emph{extends} mechanism is, of course, optional: Developers can also choose to create complete definitions from scratch, or by copying other definitions in their entirety.)

Finally, DDTS' build, run and suite definitions imply a set of dependencies that determine the execution order of test suite activities. Each run associates with a build via its \emph{build} key, making the build the run's prerequisite. Several runs may depend on the same build, so that all must acquire that build before proceeding to execution. Similarly, at the suite level, each comparison group depends on the output created by its member runs, and the suite as a whole depends on the success/failure status of each of its comparison groups. So, runs consume the product (executable programs) of builds, comparison groups consume the product (output files) of runs, and suites consume the products (results of equivalence tests) of comparison groups. End users need not concern themselves with how or when each of these requirements, established by DDTS' logic component (definition files), will be satisfied: Those details are left to the control component -- the library routines and core driver.

A separation of concerns similar to DDTS' three-complexity-level breakdown is described in \cite{rest}, where \emph{software developers} create general but more complex test components, which can then be combined and applied by \emph{software testers} to create specific test-case instances, corresponding to DDTS runs.

\subsection{Implementation: Parallelism should be used for reasonable turnaround times.}

The relationships described by DDTS' build, run and suite definition files can be viewed as a directed graph whose nodes are build, run, comparison and suite-summary activities, and whose edges reflect dependencies. In Figure \ref{figure:1}, for example, \emph{suite1} depends on comparison groups \emph{group1} and \emph{group2}; \emph{group1} depends on runs \emph{run1} and \emph{run2}; and \emph{run1} and \emph{run2} share a dependence on \emph{build1}. The two builds can execute concurrently, as can all four runs, once their build dependencies are satisfied. DDTS executes each node when all its prerequisite neighbors have finished executing. Independent activities are overlapped, via thread-based parallelism, to optimize test-suite turnaround time. Using a delegation-of-responsibility approach, each run, comparison group and the main test suite itself correspond to a single thread each. The main test-suite thread creates threads for each comparison group, joins each when it completes, and deems the entire test suite a success if each comparison group reports success within its group. Each comparison group thread, in turn, creates threads for each member run, joins them, locates each run's output, and performs comparisons between pairs of output files, reporting success or failure to the main test-suite thread based on these comparisons.

The final test-suite result is therefore built bottom-up: runs upon builds, comparisons upon runs, and the suite itself upon comparisons.

While several runs may depend on the same build, the build needs to be performed only once, and the executables shared. DDTS run threads execute builds in a mutex-protected critical region, whose first action is to check whether the build has already been performed and to skip the remainder of the region if it has. The first run thread to obtain the lock and enter the region finds that the build has \emph{not} yet been performed and so performs it, as if on behalf of all runs depending on that build. All other runs subsequently entering the critical region simply skip ahead and make use of the now-existing build.

Similar mutex-competition mechanisms decide which run, of a set sharing a common \emph{baseline} key, will contribute their output when generating a new baseline; which run will provision input data for all runs; and which comparison group will execute a run defined in more than one group (cases where this is a useful configuration exist).

This use of threads and locks allows the dependency graph to be executed, greedily, as quickly as possible, with simple logic in the DDTS core driver.

\section{Extensions to the Original Design}

Collaboration between NOAA and NASA has been beneficial to the development of DDTS, whose original design reflected the needs of modeling groups at NOAA. NASA's needs differed and indicated a need for a number of new features. For example:

\begin{itemize}
\item A suite-level \emph{build\_only} key was added to specify that only builds (no runs or comparisons) should be performed, the suite being deemed a success if all builds complete successfully.
\item A suite-level \emph{continue} key was added to specify that the test suite should not terminate on errors, but simply report them, and continue running to collect as much information as possible. At NOAA, where DDTS applications are used as pre-commit tests, failing early minimizes resource use and lets developers correct errors as soon as they are detected. (Experience has shown that multiple runs in a suite will usually fail, eventually, due to a common underlying bug, and that addressing that bug immediately is an efficient use of time.) At NASA, where DDTS applications are run as unattended regression tests, detecting all errors (not just the first) is valuable.
\item By default, DDTS deletes builds created by previous test-suite invocations to prevent false positives due to reuse of existing executables or libraries. A suite-level \emph{retain\_builds} key was added to suppress this behavior, speeding up total test-suite execution time when the source code for the program-under-test has not changed.
\item A run-level \emph{require} key was added to specify that a run may not begin executing before the run named by this key's value completes. This allows, for example, execution of a short run before a longer one is attempted.
\item The \emph{lib\_suite\_post} library routine was added to support post-suite activities like sending emails to staff with regression-test results. (The \emph{lib\_suite\_prep} routine was added for symmetry.) Also, the \emph{env} structure made available to \emph{lib\_suite\_post} was augmented with access to build and run objects from the entire test suite, for analysis and reporting purposes.
\item While the DDTS-tested NOAA models were developed with bitwise-identical output between certain configurations (e.g. different MPI task counts), and from run to run with identical configurations, as a requirement, the default bitwise-identity comparator is not universally appropriate. For instance, tests of NetCDF files created by many models may fail simply due to embedded timestamps that vary from run to run. So the ability to specify a custom comparator by defining a \emph{lib\_comp} library routine was added to support testing a wider range of output file types.
\item Originally, DDTS decided whether a run completed successfully by searching for a sentinel string, provided in the run definition, in the output messages emitted by the program-under-test. For greater flexibility, this mechanism was replaced with the \emph{lib\_run\_check} library routine, whose boolean return value indicates success or failure, and whose implementation is up to the library implementer: For example, one may need to consider the contents of multiple files to judge a run's completion status.
\item DDTS initially expected to find \emph{library.rb}, the directories containing build/run/suite definition files and related artifacts in the same directory as the DDTS core driver code. To support multiple applications with a single copy of DDTS, and different IT deployment schemes, DDTS was extended to allow overriding default paths via environment variables specifying the directories containing the core driver, the directory containing the application, and the directory where framework output (log and temporary files) should be created.
\end{itemize}

\section{Experiences}

Ruby has proven an excellent choice for DDTS development, as it has been easy to add new features, especially those relying on dynamic invocation of routines based on run-time information as, for example, when overriding generic library routines with specialized implementations (e.g. overriding \emph{lib\_run} with \emph{lib\_run\_sge} for a suite running on a platform using the SGE batch system). JRuby has provided the hoped-for portability and Ruby-version stability.

Early in the NOAA/NASA collaboration, one of the authors had the opportunity to implement a simple prototype test-suite application for a small NASA model. Although DDTS was well understood, the model itself was completely unfamiliar. It was nevertheless possible to code the program and platform interfaces, via the required routines in \emph{library.rb}, and to define a suite comprising several runs and a single build, in only a few hours. While many applications will be more complex, this exercise illustrated the ease with which an initial DDTS test-suite application, suitable for extension, can be created.

The following two subsections describe actively-used applications of DDTS to models at both NOAA and NASA.

\subsection{Applications at NOAA}

Pre-commit tests were in regular use by developers of NOAA's Flow-following finite-volume Icosahedral Model (FIM) \cite{fim} in 2011, when DDTS was initially developed, but the limitations of the imperative approach used for those tests had become clear: Each new test added to the suite meant writing new (or, worse, duplicated) code in the test scripts. Creating suites exercising different groups of runs led to increasingly complicated sets of logical conditionals to enable or disable certain runs. And determining which builds needed to be performed to meet runs' requirements, or deciding which tasks could run concurrently, was a manual and error-prone process.

The first version of DDTS was written for use with the Non-hydrostatic Icosahedral Model (NIM) \cite{nim} in 2011. In 2012, DDTS was applied to FIM as well. Since these two numerical weather-prediction models share a number of developers and general design characteristics, DDTS was well suited to both. In 2013, DDTS was applied to the Ionosphere Plasmasphere Electrodynamics (IPE) model \cite{ipe}, a space-weather prediction model also developed at NOAA. Although IPE's build, configuration and run mechanisms had much less in common with those of NIM and FIM, it proved straightforward to slot its command-line build/run utilities and manual actions into the DDTS framework.

The DDTS applications defined for these models handle a great deal of test complexity. The NIM application, for example, tests the model on three HPC platforms, using serial, MPI distributed-memory parallel and OpenMP shared-memory parallel builds; two physics packages; single- and double-precision builds; CPU and GPU architectures; and an optional parallel IO package, among many other configurations.

The nearly 40 developers of the three NOAA models use DDTS-based testing as a pre-commit check: That is, they only commit code changes to the revision-control repository when they are able to pass their respective test suites, using run-vs-run comparisons, run-vs-baseline, or both, as appropriate. Since these models' test suites are designed to be run by developers, DDTS, the associated library implementations, and each application's build/run/suite definition files are kept in the repository alongside the model code itself, so that developers always check out everything together. This arrangement maintains a rigorous relationship between the current model code and the test suites appropriate to it.

While no formal study has been performed on the effects of pre-commit testing on these development efforts, it can be confidently, albeit anecdotally, reported that test suites fail often, and that the underlying errors so detected would be dangerous, confusing, difficult to debug, and a massive waste of time were they to be committed to revision control and foisted on other developers. Bugs are easiest to fix when the changes causing them are fresh in the mind -- especially in the mind of the developer who introduced them.

\subsection{Applications at NASA}

Several NASA science codes, which were already undergoing systematic regression testing using various ad hoc solutions, standardized on DDTS for system testing. This decision led to greater flexibility in test design and lower maintenance effort due to common infrastructure.

DDTS' simple workflow was found to be sufficient to support regression-testing efforts at NASA Goddard Space Flight Center and, extended as described in the previous section, is flexible enough to handle idiosyncrasies of the several tested models. For example, enabling the suite-level \emph{continue} feature allows DDTS to to attempt execution of all runs in a suite, even if one or more fails. This behavior differs from the ``fail fast'' mode used for pre-commit testing by NOAA developers.

While NOAA's DDTS applications are packaged with the model code they test, this relationship is reversed at NASA: DDTS is maintained separately, with custom applications for each tested model. Each application's \emph{lib\_build\_prep} routine polls the revision-control repository for changes to model code, and time-consuming updates and builds are performed only if the model has changed (a behavior made possible by the suite-level \emph{retain\_builds} setting). In addition, a set of routines used across multiple applications, callable from the standard library routines, is maintained separate from \emph{library.rb}.

The NASA-Unified Weather Research and Forecasting (NU-WRF) \cite{nuwrf} model consists of several executable pre-processors that must run in a specific sequence to attain the desired output. The NU-WRF workflow has been defined as a DDTS run, parameterized via run-definition files. The configuration information exposed in YAML, controlling Ruby code in library routines, is sufficient to generate a large variety of tests without further programming, allowing greatly increased coverage of tested preprocessors.

The Land Data Toolkit (LDT), which can generate input data for NU-WRF simulations, is currently undergoing rapid development as part of the GSFC Land Information System (LIS) \cite{lis} and will be published with LIS version 7. LDT testing requires exercising a large number of run-time options, achieved mainly by making small changes to the input test data, to increase internal code coverage. This use case leverages DDTS' \emph{extends} mechanism to derive a large number of similar run definitions from a common ancestor, differentiating runs by overriding a small number of the inherited settings.

DDTS is also under consideration for long-term use with the NASA Goddard Institute for Space Studies modelE project \cite{modele}, and preliminary results indicate that it is possible to break up modelE's build and run phases to fit the DDTS framework. DDTS' run-level \emph{require} mechanism is especially useful for modelE tests, as it makes it possible to attempt a long simulation only after a short one has run successfully. This is an important HPC resource-management consideration, giving up some available parallelism to avoid wasting valuable compute resources. DDTS' \emph{require} mechanism is also used to test ``restart'' runs by using the output of a completed run as the input to another.

NASA's success in transitioning testing for these models indicates that adopting DDTS can be straightforward and advantageous even in cases where some form of testing is already in place.

\subsection{Other Applications}

One of the authors has also created and demonstrated prototype DDTS applications for both the NOAA Environmental Modeling System (NEMS) \cite{nems}, a multi-model numerical weather prediction framework being developed at the U.S. National Centers for Environmental Prediction; and for Yonsei University's Global/Regional Integrated Model system (GRIMs) \cite{grims}. Both NEMS and GRIMs exhibited design and structural elements different from the models previously tested with DDTS, but proved amenable to testing with the framework.

\subsection{Summary}

\begin{table}[!t]
\centering
\begin{tabular}{r c c c c c c}
\hline
& FIM & IPE & LIS LDT & modelE & NIM & NU-WRF \\ 
\hline
Platforms Supported & 2 & 1 & 1 & 1 & 3 & 1\textsuperscript{3} \\
Suite Definitions & 9 & 1 & 2 & 4 & 3 & 18 \\
Run Definitions & 56 & 7 & 36 & 21 & 122 & 50 \\
Build Definitions & 16 & 4 & 2 & 40 & 27 & 32\textsuperscript{4} \\
DDTS LoC\textsuperscript{1} & 1112 & 852\textsuperscript{2} & 1112 & 1112 & 1112 & 1112 \\
Library LoC\textsuperscript{1} & 487 & 164 & 931 & 876 & 381 & 1500 \\
Definitions LoC\textsuperscript{1} & 674 & 70 & 549 & 495 & 1164 & 4171 \\
\hline
\newline
\end{tabular}
1. LoC=Lines of Code. 2. Older version. 3. Support for a second HPC platform is planned. 4. Not all in use currently.
\newline
\caption{Metrics for models currently tested with DDTS.}
\label{table:metrics}
\end{table}

Table \ref{table:metrics} shows metrics, including counts of supported platforms, build/run/suite definitions, and lines of code, for several scientific codes using DDTS. Note that not every definition file necessarily represents a complete build, run or suite: Some are merely fragments establishing common settings, meant to be extended in other definition files.

\section{Future Work}

Future work may address the following current limitations:

\begin{itemize}
\item When using DDTS' standard bitwise-exact comparator, a relatively small number of tests is sufficient, due to the transitive property of equality. DDTS may designate one run the ``master'' and compare each other run against it: If each run is bitwise-identical to the master, all runs are bitwise identical. When using an alternative comparator (e.g. equality within some defined tolerance) however, the transitive property no longer holds: The fact that a tested value from run A is equal to its counterpart from run B within tolerance, and B equal to C within the same tolerance, does not imply that the values from A and C are equal within tolerance. Therefore, DDTS should be equipped with an option to conduct a (potentially much larger) set of all-pairs comparisons when using an alternative comparator.
\item Currently, DDTS only allows a single run thread, via the mutex-locking mechanism previously described, to execute \emph{lib\_data} to obtain and prepare input data for all runs in the test suite, under the assumption that all runs have the same data requirements. This is too restrictive and should be made flexible to allow runs to associate with different input data sets, in the same way they associate with different baselines.
\item The \emph{require} key should be extended to support builds as well as runs, which would allow finer-grained tests of complex build systems, as well as builds of supported libraries.
\item The single-inheritance \emph{extends} mechanism may also be too restrictive: Definition ancestry chains can quickly grow long, and designing a robust hierarchy may need careful planning. A ``mixin''-style \emph{include} mechanism that would allow insertion of arbitrary definition fragments anywhere in the hierarchy might be helpful.
\end{itemize}

\section{Conclusion}

DDTS is a practical testing framework designed to introduce routine testing into scientific-software projects as painlessly as possible. It has been used to implement both developer pre-commit test suites and unattended regression test suites, and has undergone active development to respond the needs of real-world scientific-software projects at NASA and NOAA, supporting teams ranging from a small handful of developers to several dozen, and accommodating various deployment environments. It leverages the Ruby programming language and the JRuby implementation to remain lightweight and portable, and YAML to provide a simple, declarative domain-specific language to define test-suite activities and the dependency relationships between them.

The DDTS framework's design incorporates sound software-engineering principles and exposes some of them (e.g. DRY) to the test-suite application developer. The user who ultimately invokes the framework only needs to deal with a simple command-line interface, but is welcomed to participate in the design of new test cases and suites by the simple declarative syntax of the build, run and suite definitions.

It is easy to start testing with DDTS: A small set of routines must be defined to interface DDTS to the program-under-test and the computer platform, then simple YAML written to define builds, runs, comparison groups and test suites. Iterative test-suite development is encouraged: One can explore the testing problem, building up a complete DDTS application one piece at a time, from builds up to suites, without knowing beforehand exactly what the result will look like, and employing advanced configuration options only as requirements present themselves. At its simplest, a DDTS application can organize, standardize and automate the processes developers now perform manually, introducing systematic routine testing into their projects and, hopefully, helping scientific-software development efforts proceed on a sound basis.

DDTS is open-source, and distributed under a free-software license. It may be obtained at \emph{https://github.com/maddenp/ddts}. The online repository includes a sample application that demonstrates many of the features described in this paper.

\section*{Acknowledgment}

The authors would like to thank Tom Henderson, whose original test system for FIM defined (and met) many of the DDTS design criteria, and whose feedback on the design and implementation of DDTS has been invaluable.


\begin{thebibliography}{1}

\bibitem{prags}
A. Hunt and D, Thomas, \emph{The Pragmatic Programmer: From Journeyman to Master}. Boston, MA: Addison-Wesley, 2000.

\bibitem{fowler-ci}
M. Fowler. (2006, May 1). \emph{Continuous Integration} [Online]. Available: \url{http://martinfowler.com/articles/continuousIntegration.html}

\bibitem{extreme}
D. Wells. (2013, October 8). \emph{Extreme Programming: A gentle introduction} [Online]. Available: \url{http://www.extremeprogramming.org/}

\bibitem{batlab}
A. Pavlo \emph{et al}., ``The NMI build \& test laboratory: continuous integration framework for distributed computing software,'' in \emph{Proc. 20th Conf. on Large Installation System Administration}, Berkeley, CA, 2006, pp 263-273.

\bibitem{comp-sci-survey}
P. Prabhu \emph{et al}., ``A survey of the practice of computational science,'' in ``State of the Practice Reports (SC '11)'', ACM, New York, NY, Article 19, 2011.

\bibitem{cig}
E. M. Heien \emph{et al}. (2013, September 4). \emph{Experiences with Automated Build and Test for Geodynamics Simulation Codes} [Online]. Available: \url{http://arxiv.org/abs/1309.1199}

\bibitem{does-not-compute}
Z. Merali. (2010, October 13). \emph{Computational science: ...Error ...why scientific programming does not compute} [Online]. Available: \url{http://www.nature.com/news/2010/101013/pdf/467775a.pdf}

\bibitem{bp4sc}
G. Wilson, \emph{et al}. (2013, September 26). \emph{Best Practices for Scientific Computing} [Online]. Available: \url{http://arxiv.org/abs/1210.0530}

\bibitem{how-do}
J. E. Hannay, \emph{et al}., ``How do scientists develop and use scientific software?'' in \emph{Proc. 2009 ICSE Workshop on Software Engineering for Computational Science and Engineering}, Vancouver, BC, 2009, pp 1-8.

\bibitem{bottleneck}
G. Wilson, ``Where's the real bottleneck in scientific computing?'' \emph{American Scientist}, vol. 94, no. 1, pp. 5-6, January, 2006.

\bibitem{sw-carp}
G. Wilson, ``Software Carpentry: Getting Scientists to Write Better Code by Making Them More Productive,'' \emph{IEEE Comput. Sci. Eng. Mag}, vol. 8, no. 6, pp. 66-69, November/December, 2006.

\bibitem{ssi}
S. Crouch \emph{et al}., ``The Software Sustainability Institute: Changing Research Software Attitudes and Practices,'' \emph{IEEE Comput. Sci. Eng. Mag}, vol. 15, no. 6, pp. 74-80, November/December, 2013.

\bibitem{clune}
T. L. Clune and R. B. Rood, ``Software Testing and Verification in Climate Model Development,'' \emph{IEEE Softw.}, vol. 28, no. 6, pp. 49-55, November/December, 2011.

\bibitem{junit}
\emph{JUnit} [Online]. Available: \url{http://junit.org/}

\bibitem{pfunit}
\emph{pFUnit} [Online]. Available: \url{http://pfunit.sourceforge.net/}

\bibitem{decl-advantages}
J. W. Lloyd, ``Practical Advantages of Declarative Programming,'' in \emph{Joint Conf. on Declarative Programming}, Peniscola, Spain, 1994, pp. 18-30.

\vspace{141mm} % equalize columns

\bibitem{ruby}
\emph{Ruby} [Online]. Available: \url{https://www.ruby-lang.org/}

\bibitem{jruby}
\emph{JRuby} [Online]. Available: \url{http://jruby.org/}

\bibitem{logic-control}
R. Kowalski, ``Algorithm = logic + control,'' in \emph{Communications of the ACM}, vol. 22, no. 7, pp. 424-436, July, 1979.

\bibitem{yaml}
\emph{YAML} [Online]. Available: \url{http://www.yaml.org/}

\bibitem{rest}
C. H. Kao, C. C. Lin, J. Chen, ``Performance Testing Framework for REST-Based Web Applications,'' \emph{Quality Software}, Nanjing, 2013, pp. 349-354.

\bibitem{fim}
\emph{The FIM Global Model} [Online]. Available: \url{http://fim.noaa.gov/}

\bibitem{nim}
J. Lee and A. E. McDonald. (2010, March 9). ``NIM: Nonhydrostatic Icosahedral Model'' [Online]. Available: \url{http://www.esrl.noaa.gov/research/review/2010/posters/2-7-Jin.Lee.pdf}

\bibitem{ipe}
Maruyama \emph{et al}., ``The Ionosphere-Plasmasphere-Electrodynamics (IPE) Model: An Impact of the Realistic Geomagnetic Field Model on the Ionospheric dynamics and energetics,'' to be submitted to \emph{J. Geophys. Res.}, 2014a.

\bibitem{nuwrf}
\emph{NASA-Unified Weather Research and Forecasting (NU-WRF) model} [Online]. Available: \url{https://modelingguru.nasa.gov/community/atmospheric/nuwrf}

\bibitem{lis}
\emph{Land Information System (LIS)} [Online]. Available: \url{http://lis.gsfc.nasa.gov/LIS_whatis.php}

\bibitem{modele}
\emph{GISS GCM ModelE} [Online]. Available: \url{http://www.giss.nasa.gov/tools/modelE/}

\bibitem{nems}
\emph{NEMS} [Online]. Available: \url{https://earthsystemcog.org/doc/detail/1850/}

\bibitem{grims}
\emph{GRIMs} [Online]. Available: \url{https://www.grims-model.org/start.jsp}


\end{thebibliography}
\end{document}